\documentclass[aps,prd,twocolumn,superscriptaddress]{revtex4}
\usepackage{graphicx}
\begin{document}

% Use the \preprint command to place your local institutional report
% number in the upper righthand corner of the title page in preprint mode.
% Multiple \preprint commands are allowed.
% Use the 'preprintnumbers' class option to override journal defaults
% to display numbers if necessary
%\preprint{}

%Title of paper
\title{Newtonian Perturbations on Models with Matter Creation}

\author{J. F. Jesus}
\email[]{jfernando@astro.iag.usp.br}
\author{F. A. Oliveira}
\email[]{foliveira@astro.iag.usp.br}
\affiliation{Instituto de Astronomia, Geof\'{\i}sica e Ci\^encias Atmosf\'ericas, Universidade de S\~ao Paulo,\\
Rua do Mat\~ao, 1226 - Cidade Universit\'aria, 05508-080, S\~ao
Paulo, SP, Brasil}
\author{S. Basilakos}
\email[]{svasil@academyofathens.gr} \affiliation{Academy of Athens,
Research Center for Astronomy and Applied Mathematics, Soranou
Efesiou 4, 11527, Athens, Greece}
\affiliation{High Energy Physics Group, Dept. ECM, Universitat de Barcelona,
Av. Diagonal 647, E-08028 Barcelona, Spain}
\author{J. A. S. Lima}
\email[]{limajas@astro.iag.usp.br}

\affiliation{Instituto de Astronomia, Geof\'{\i}sica e Ci\^encias Atmosf\'ericas, Universidade de S\~ao Paulo,\\
Rua do Mat\~ao, 1226 - Cidade Universit\'aria, 05508-900, S\~ao
Paulo, SP, Brasil}

\date{\today}

\begin{abstract}
Creation of Cold Dark Matter (CCDM) can  macroscopically be
described by a negative pressure, and, therefore, the mechanism is
capable to accelerate the Universe, without the need of an
additional dark energy component. In this framework we discuss the
evolution of perturbations by considering a Neo-Newtonian approach
where, unlike in the standard Newtonian cosmology, the fluid
pressure is taken into account even in the homogeneous and isotropic
background equations (Lima, Zanchin and Brandenberger, MNRAS {\bf
291}, L1, 1997). The evolution of the density contrast is calculated
in the linear approximation and compared to the one predicted by the
$\Lambda$CDM model. The difference between the CCDM and $\Lambda$CDM
predictions at the perturbative level is quantified by using three
different statistical methods, namely:  a simple $\chi^{2}$-analysis
in the relevant space parameter,  a Bayesian statistical inference,
and, finally, a Kolmogorov-Smirnov test. We find that under certain
circumstances the  CCDM scenario analyzed here predicts an overall
dynamics (including Hubble flow and matter fluctuation field) which
fully recovers that of the traditional cosmic concordance model. Our
basic conclusion is that such a reduction of the dark sector
provides a viable alternative description to the accelerating
$\Lambda$CDM cosmology.
\end{abstract}


\maketitle

\section{\label{sec1}Introduction}
The growing evidences for cosmic acceleration as, e.g., the
Supernova Ia observations \cite{riess98,Ko2008,Hi2009,Aman2010}, is
challenging cosmologists and theoretical physicists. The neediness
of some new ingredient in  the cosmic recipe, in order to preserve
Einstein's Equations, inspired both communities to conservatively
invoking the simplest available hypothesis, namely, a cosmological
constant, $\Lambda$.

Nevertheless, the identification of $\Lambda$ with the quantum
vacuum has brought another problem: the estimate of theoretical
physicists that the vacuum energy density should be 120 orders of
magnitude bigger than the measured $\Lambda$ value. This is the
``old" cosmological constant problem (CCP) \cite{weinb89}. The
``new" problem \cite{Peebles03} asks why is the vacuum density so
similar to the matter density just now? Many solutions to both
problems have been proposed in the literature
\cite{Zlatev99,AASW09}. The majority of them are plagued with no
physical basis and/or many parameters.

On the other hand, some authors have also investigated a class of
models where the creation of cold dark matter may result on a
pressure which is negative (named CCDM, Creation of Cold Dark
Matter), thereby providing a mechanism for cosmic acceleration.
In the current literature, the leitmotiv of such models is to
reduce the dark sector (dark energy + dark matter) in 
the framework of general relativity since
the existence of dark energy seems to be less secure 
than dark matter \cite{BDRS03,PDM}.
The extra bonus is to solve the coincidence and cosmological 
constant problems.

The search for dark matter accelerating 
models started almost one decade before the
SNe Ia observations. Initially, Prigogine and coworkers \cite{Pri89}
argued that the gravitationally-induced particle creation could
consistently be discussed in the realm of the relativistic
non-equilibrium thermodynamics. Later on, their macroscopic
formulation was clarified by Calv\~ao, Lima and Waga through a
manifestly covariant formulation \cite{CLW92}. The inclusion of the
back reaction in the Einstein Field Equations (EFE) via an effective
pressure (which is negative for an expanding space-time) opened the
way for cosmological applications. As a consequence, several
interesting features of cosmologies where the dark sector is reduced
due to the creation of CDM matter have been discussed in the last
decade \cite{LSS08,SSL09}.

More recently, a new CCDM cosmology was proposed by Lima, Jesus and
Oliveira (from now on LJO model) which mimics the Hubble expansion
history of the $\Lambda$CDM model, at least at the level of the
background equations \cite{LJO}.

The quoted LJO model is quite interesting, once that it perfectly
mimics the $\Lambda$CDM  cosmic history, thereby also  providing a
good fit to current cosmological data (SNIa, BAO and CMB shift
parameter, total age, etc) with the same number of parameters of the
$\Lambda$CDM, but without the CCP. In this sort of models, the
quantum vacuum energy is canceled out by some not yet known physical
mechanism. A basic advantage of this kind of scenario comes from the
fact that a vacuum energy density which is null or negligible is
more acceptable than a vacuum energy density finely tuned like in
the $\Lambda$CDM model.

Naturally, although providing good fits to background data, it would
be interesting to investigate to which level this similarity is
preserved, for instance, by analyzing  the evolution of small
fluctuations predicted by the CCDM equations. Preliminary studies on
this subject has already been accomplished by Basilakos and Lima
\cite{BasLim}, however, focusing on the theoretical consequences to
cluster abundances at different redshifts  (see also
\cite{BPL2010}). By using the Press-Schechter formalism, they also
investigated  the cluster-size halo redshift distribution by
confronting the results with future cluster surveys (eROSITA
satellite and Sunayev-Zeldovich survey based on the South Pole
Telescope). We would like to stress that in the latter papers 
we ignored possible contributions from pressure at 
the perturbative level.

In the present paper, we are basically interested to determine
whether the LJO model provides a realistic description at the
perturbative level, however, by fitting the available fluctuation
data, like the growth rate of clustering. The difference between the
CCDM and $\Lambda$CDM predictions at the perturbative level is
quantified using tree different statistical methods, namely: a
simple $\chi^{2}$-analysis in the relevant space parameter, a
Bayesian statistical inference, and, finally, a Kolmogorov-Smirnov
test. As we shall see, the CCDM reduction dark sector provides a
viable alternative description to the accelerating $\Lambda$CDM
cosmology.

The work is planned as follows. The general relativistic approach
for CCDM models is presented in section II. In sections III we show
how the Neo-Newtonian treatment adopted here recovers the
relativistic formulation whereas, in  section IV, we derive the
associated evolution equation for the density contrast. In section
V,  we discuss in detail the linear growth factor of matter
perturbations, the CCDM theoretical predictions regarding the
evolution of the growth rate of clustering, and the corresponding
statistical analyzes. The main conclusions are summarized in VI.

\section{CCDM Cosmology in the matter era: Relativistic Formulation}
The background cosmological equations of the model with creation of
cold dark matter (CCDM) have  the following form (for simplicity, we
are neglecting the contributions of the radiation and baryonic
components):
\begin{eqnarray}
\label{FRW} 8\pi G\rho&=&3\frac{\dot{a}^2}{a^2}+3\frac{K}{a^2},\\
\label{FRW2} 8\pi
Gp_c&=&-2\frac{\ddot{a}}{a}-\frac{\dot{a}^2}{a^2}-\frac{K}{a^2}.
\end{eqnarray}

\noindent where $\rho$ is the CDM density, $p_c$ is the creation
pressure,  $a(t)$ is the scale factor, and an overdot means time
derivative. In the case of constant specific entropy (per particle),
the creation pressure for a CDM component is given by
\cite{Pri89,CLW92}

\begin{equation}
\label{Eqp_c} p_c=-\frac{\rho\Gamma}{3H},
\end{equation}

\noindent where $\Gamma$ is the creation rate of CDM particles and
$H=\dot{a}/a$ is the Hubble parameter.

It is readily checked that equations (\ref{FRW}) and (\ref{FRW2})
lead  to the continuity and acceleration equations:

\begin{eqnarray}
\label{ContCCDM}
\frac{\dot{\rho}}{\rho}+3\frac{\dot{a}}{a}&=&\Gamma,\\
\label{AccelCCDM} \frac{\ddot{a}}{a}=-\frac{4\pi
G}{3}(\rho+3p_c)&=&-\frac{4\pi
G\rho}{3}\left(1-\frac{\Gamma}{H}\right).
\end{eqnarray}

The above Eqs. (\ref{ContCCDM}-\ref{AccelCCDM}) imply that the model
is fully determined by the creation rate $\Gamma$, or more
precisely, by the ratio $\Gamma/H$. If $\Gamma << H$, the particle
creation process is negligible, and the standard dust filled models
are recovered. In the LJO scenario, the phenomenological creation
rate of cold dark matter has been defined by:
\begin{equation}
\label{Gamma}
\frac{\Gamma}{H} =3\tilde{\Omega}_{\Lambda} \frac{\rho_{c0}}{\rho} ,
\end{equation}
where  $\tilde{\Omega}_{\Lambda}$ (termed $\alpha$ in the original
LJO notation \cite{LJO}) is a constant parameter, ${\rho_{c0}}$ is
the present day value of the critical density, and the factor 3 has
been added for mathematical convenience.

Now, by inserting the above expression into the energy conservation
as given by (\ref{ContCCDM}) one obtains
\begin{equation}
\dot{\rho} + 3H\rho = \Gamma \rho \equiv 3\tilde{\Omega}_{\Lambda}
{\rho_{c0}}H,
\end{equation}
which can be readily integrated to give a solution for the energy
density

\begin{equation}
\label{rhodm}
\rho = (\rho_{0} - \tilde{\Omega}_{\Lambda}\rho_{c0})a^{-3} +
\tilde{\Omega}_{\Lambda}\rho_{c0}.
\end{equation}

In terms of the cosmic history, the equivalence between this model
and the $\Lambda$CDM model can be seen directly through the
evolution equation of the scale factor function.  As one may check,
by inserting the expression of the creation pressure $p_c$ in Eq.
(\ref{FRW2}) we obtain:
\begin{equation}
\label{evola}
2a{\ddot a}+ {\dot{a}}^2 + K - 3\tilde{\Omega}_{\Lambda}H_0^2 a^{2}
= 0,
\end{equation}
which should be compared to:
\begin{equation}
\label{evolaLCDM} 2a{\ddot a}+ {\dot{a}}^2 + K - {\Lambda}a^{2}  =
0,
\end{equation}
provided by the $\Lambda$CDM model. The above equations imply that
the models will have the same dynamical behavior when we identify
the creation parameter by the expression
$\tilde{\Omega}_{\Lambda}={\Lambda}/3{H_0}^{2}
\equiv\Omega_{\Lambda}$.
Further, the factor 
$\tilde{\Omega}_{m} \equiv \Omega_m - \tilde{\Omega}_{\Lambda}$,
where $\Omega_m\equiv\frac{\rho_0}{\rho_{c0}}$, can  also be
identified as an `effective' CDM density parameter, in comparison to
the $\Lambda$CDM model \cite{LJO}.

In the context of a spatially flat geometry ($K=0$ and
$\Omega_m=1$), one may show that the inflection point, that is, the
point where the Hubble expansion changes from the decelerating to
the accelerating regime  ($\ddot{a}(t_{I})=0$) takes place at:
\begin{eqnarray}
     \label{infle}
a_{I}=\left[\frac{\tilde
{\Omega}_{m}}{2\tilde{\Omega}_{\Lambda}}\right]^{1/3} \;.
\end{eqnarray}
As an example for
$\tilde{\Omega}_{m}=1-\tilde{\Omega}_{\Lambda}=0.30$ we find
$a_{I}\simeq 0.60$ (or $z_I\simeq 0.67$). Finally, the necessary
condition for an inflection point in our past is $a_{I}<1$, which
leads to the condition $\frac{1}{3}<\tilde{\Omega}_{\Lambda}<1$.

\section{Neo-Newtonian formulation for CCDM Cosmologies}
The relativistic dynamics of CCDM cosmologies is the same of a
single fluid with density $\rho$ and pressure $p_c$ (see Eqs.
(\ref{FRW}-\ref{Eqp_c})). In addition, whether the equivalent fluid
has an equation of state (hereafter EoS) parameter, $w=p/\rho$,
taking a look at Eq. (\ref{Eqp_c}), we may see that this fluid
($p=p_c$), in general, has a time varying equation of state
parameter, $w=-\frac{\Gamma}{3H}$.

In this connection, it is interesting to notice that fluids with
non-vanishing pressure were also studied by Lima, Zanchin and
Brandenberger \cite{lima97} within a `quasi'-Newtonian perspective.
Following earlier ideas developed by McCrea \cite{MAC51} and
Harrison \cite{H65}, they proposed the so called Neo-Newtonian
description in order to reproduce the relativistic equations with
pressure either at the background and perturbative levels. The basic
equations of such a description are given by:
\begin{equation}
\label{Euler} \left(\frac{\partial{\bf u}}{\partial t}\right)_r
+({\bf u}\cdot\nabla_r){\bf u}=- \nabla_r\Phi-
\frac{\nabla_rp_c}{\rho+p_c}
\end{equation}
\begin{equation}
\label{Continuity} \left(\frac{\partial\rho}{\partial
t}\right)_r+\nabla_r\cdot(\rho{\bf u})+p_c\nabla_r\cdot{\bf u}=0
\end{equation}
\begin{equation}
\label{Poisson} \nabla_r^2\Phi=4\pi G(\rho+3p_c)
\end{equation}
where we have already included the creation pressure, $p_c$. These
equations are named Euler, continuity and Poisson equations,
respectively. The modified continuity equation is due to Lima {\it
et al.} \cite{lima97}, in order to account correctly to pressure
effects. As one may check, such equations reproduce the FRW type
equations with pressure in the homogeneous and isotropic case
($p_c=p_c(t)$, $\rho=\rho(t)$), and can also be consistently
perturbed for any given equation of state (see discussion below). In
particular, they also shown that the linear perturbed version of
these equations yield the correct growing mode for the density
contrast of a fluid with $p=w\rho$, with constant $w$. For the case
that $w$ is a time dependent quantity, the linear perturbation is
equivalent to the full general relativistic formulation, at least
when some conditions are imposed \cite{Reis2003}. Indeed,
applications of this Neo-Newtonian approximation are not restricted
to nonrelativistic matter, and the high accuracy of the
approximation has also been proved  for different epochs and even
for scales larger than Hubble radius, as well as, for the spherical
gravitational collapse \cite{Pad98,abramo07}. Therefore, it is
reasonable to expect that the Neo-Newtonian approach will provide a
good approximation for the evolution of  small density fluctuations
in the CCDM case.

\subsection{Recovering the Relativistic CCDM Model from the Neo-Newtonian Approach}

Let us now discuss how the basic relativistic equations of CCDM
model can be recovered by the set of equations (\ref{Euler}),
(\ref{Continuity}), and (\ref{Poisson}). In order to show that, we
first remark that homogeneity and spatial isotropy of the
unperturbed model implies $\mathbf{u}= H \mathbf{r}$, according to
the Hubble's law. Then, Eq. (\ref{Continuity}) can be rewritten as:

\begin{equation}
\dot{\rho} + 3H\rho = \Gamma \rho.
\end{equation}
Assuming the creation rate $\Gamma$ as in the Eq.(\ref{Gamma}), the
above equation can readily be integrated, to give exactly the same
expression for the density evolution in Eq.(\ref{rhodm}). The Euler
equation (\ref{Euler}) can be rewritten as
\begin{equation}
\label{euler2} \dot{H} {\bf r} + H r \frac{\partial}{\partial r}(H
{\bf r})= \left( \dot{H} + H^2 \right)  {\bf r} = - {\bf \nabla}_r
\Phi.
\end{equation}
While the integration of the Poisson equation (\ref{Poisson})
yields:
\begin{equation}
\label{poisson2} {\bf \nabla}_r \Phi = 4 \pi G (\rho + 3p_c) \frac{
{\bf r}}{3},
\end{equation}

Finally, by combining Eqs. (\ref{euler2}) and (\ref{poisson2}), we
derive the acceleration equation:
\begin{equation}
\label{accel2} \frac{\ddot{a}}{a}=-\frac{4 \pi G}{3}(\rho + 3p_c),
\end{equation}

\noindent where, as in the relativistic case, there is the presence
of the creation pressure term, which can give rise for acceleration
since it is negative for the expanding Universe. Using the
expressions (\ref{Eqp_c}) and (\ref{rhodm}), the equation above can
be integrated resulting in the following expression:

\begin{equation}
\label{fried4} 8 \pi G \rho= 3 H^ 2 + \frac{3K}{a^ 2}.
\end{equation}

\noindent where $K$ is an arbitrary integration constant. This is
Friedmann equation (\ref{FRW}).

Thus,  Eqs. (\ref{accel2}) and (\ref{fried4}) show that the
background cosmological equations can effectively be recovered by
the Neo-Newtonian formulation. This result is valid regardless of
the specific form assumed to the creation rate $\Gamma$, and should
be compared with the incomplete treatment adopted by Roany and
Pacheco \cite{RP10}, and,  critically, rediscussed by Lima et al.
\cite{LJO2}.

\section{\label{sec2}Neo-Newtonian Density Perturbations}
In general, a perturbative analysis in cosmology requires a full
relativistic description since the standard non-relativistic
(Newtonian) approach works well only when the scale of perturbation
is much less than the Hubble radius and the velocity of peculiar
motions are small in comparison to the Hubble flow \cite{peeb80}.
However, as remarked earlier, such difficulties have been
circumvented by the Neo-Newtonian approximation adopted here, and,
therefore, such an approach may provide a good feeling about the
behavior of the density perturbations.

To begin with, let us transform the Neo-Newtonian Equations
(\ref{Euler})-(\ref{Poisson}) to comoving coordinates ${\bf x}$,
which are related to the proper coordinates ${\bf r}$ by
\cite{peeb93,lima97}:
\begin{equation}
{\bf x}=\frac{{\bf r}}{a}. \label{eq:8}
\end{equation}
In these coordinates, the basic quantities ${\bf u}$, $\rho$ and
$\Phi$ can be rewritten as:
\begin{equation}
{\bf u}=\dot{a}{\bf x}+a\dot{{\bf x}}=\dot{a}{\bf x}+{\bf v},
\label{eq:9}
\end{equation}
\begin{equation}
\rho=\bar{\rho}(t)[1+\delta({\bf x},t)], \label{eq:10}
\end{equation}
\begin{equation}
p_c=\bar{p}_c(t)+\delta p_c({\bf x},t),
\end{equation}
\begin{equation}
\Phi=\phi+\frac{2\pi Ga^2}{3}(\bar{\rho}+3\bar{p}_c)x^2,
\label{eq:12}
\end{equation}
where ${\bf v}$ is the velocity perturbation (peculiar velocity),
$\delta$ is the dark matter density contrast, $\delta p_c$ is the
perturbation of creation pressure and $\phi$ is the peculiar
gravitational potential, which generates the peculiar acceleration.
Here, variables with an overbar represent the background quantities.

Following standard lines, we  change the proper coordinates (${\bf
r}$, $t$) to comoving coordinates (${\bf x}$, $t$) and using the
differential operator identities
\begin{equation}
\nabla_x\equiv\nabla=a\nabla_r, \label{eq:13}
\end{equation}
and
\begin{equation}
\left(\frac{\partial}{\partial
t}\right)_x\equiv\frac{\partial}{\partial t} =
\left(\frac{\partial}{\partial t}\right)_r+\frac{\dot{a}}{a}{\bf
x}\cdot\nabla_x, \label{eq:14}
\end{equation}
Eqs. (\ref{Euler})-(\ref{Poisson}) can be rewritten as:

\begin{equation}
\frac{\partial{\bf v}}{\partial t}+\frac{\dot{a}}{a}{\bf
v}+\frac{({\bf v}\cdot\nabla){\bf v}}{a}=
-\frac{\nabla\phi}{a}-\frac{\nabla \delta p_c}{a(\rho+p_c)},
\label{eq:15}
\end{equation}

\begin{equation}
\bar{\rho}\frac{\partial\delta}{\partial t} + 3H(\delta
p_c-\bar{p}_c\delta) + \frac{\rho+ p_c}{a}\nabla\cdot{\bf
v}+\frac{\bar{\rho}}{a}{\bf v}\cdot\nabla\delta=0, \label{eq:16}
\end{equation}

\begin{equation}
\nabla^2\phi=4\pi Ga^2(\bar{\rho}\delta+3\delta p_c). \label{eq:17}
\end{equation}
It should be stressed that the above
Eqs.~(\ref{eq:15})-(\ref{eq:17}) are obtained only by using  the
unperturbed background Eqs. (\ref{Eqp_c})-(\ref{AccelCCDM}). In
order to simplify the above equations it is suitable to define the
quantities:
\begin{equation}
w\equiv\frac{\bar{p}_c}{\bar{\rho}} \;\;\;\;\;
c_{eff}^2\equiv\frac{\delta p_c}{\delta\rho}=\frac{\delta
p_c}{\bar{\rho}\delta}. \label{eq:wc}
\end{equation}
Now,  Eqs.~(\ref{eq:15})-(\ref{eq:17}) can be rewritten as:
\begin{equation}
\frac{\partial{\bf v}}{\partial t}+H{\bf v}+\frac{({\bf
v}\cdot\nabla){\bf v}}{a}=
-\frac{\nabla\phi}{a}-\frac{\nabla(c_{eff}^2\delta)}{a[1+w+(1+c_{eff}^2)\delta]},
\label{eq:19}
\end{equation}
\begin{equation}
\frac{\partial\delta}{\partial t} + 3H(c_{eff}^2-w)\delta +
\frac{1+w+(1+c_{eff}^2)\delta}{a}\nabla\cdot{\bf v}+\frac{{\bf
v}\cdot\nabla\delta}{a}=0, \label{eq:20}
\end{equation}
\begin{equation}
\nabla^2\phi=4\pi Ga^2\bar{\rho}\delta(1+3c_{eff}^2). \label{eq:21}
\end{equation}
These equations can already be compared to the equations of
\cite{abramo07}, where they study the perturbations of a mixture of
non-interacting fluids using various EoS parameters.
These are identical to their
equations for the case of one fluid with 
an EoS parameter $w$.

Here we are interested only on the linear order of the
perturbations. In this case,  we see that Eq. (\ref{eq:21}) is
already linear  while Eqs. (\ref{eq:19})-(\ref{eq:20}) are reduced
to:
\begin{equation}
\frac{\partial{\bf v}}{\partial t}+H{\bf v}=
-\frac{\nabla\phi}{a}-\frac{\nabla(c_{eff}^2\delta)}{a(1+w)},
\label{eq:22}
\end{equation}
\begin{equation}
\frac{\partial\delta}{\partial t} + 3H(c_{eff}^2-w)\delta +
\frac{1+w}{a}\nabla\cdot{\bf v}=0. \label{eq:23}
\end{equation}
Now, by calculating the divergent of Eq. (\ref{eq:22}) and inserting
(\ref{eq:21}) in the resulting equation, it is easy to see that Eq.
(\ref{eq:22}) takes the form
\begin{equation}
\frac{\partial\theta}{\partial t}+H\theta= -4\pi
Ga\bar{\rho}\delta(1+3c_{eff}^2) +
\frac{k^2c_{eff}^2\delta}{a(1+w)}, \label{eq:24}
\end{equation}
where we have defined $\theta\equiv\nabla\cdot{\bf v}$ and assumed
$c_{eff}^2=c_{eff}^2(t)$ and that the spatial dependence of $\delta$
is proportional to $e^{i{\bf k}\cdot{\bf x}}$. Now, we may isolate
$\theta=\nabla\cdot{\bf v}$ on (\ref{eq:23}), and replace it into
(\ref{eq:24}) to finally find:

\begin{widetext}
\begin{eqnarray}
\frac{\partial^2\delta}{\partial t^2} + \left[H(2+3c_{eff}^2-3w) -\frac{\dot{w}}{1+w}\right]\frac{\partial\delta}{\partial t}&+& \nonumber\\
+ \left\{3(\dot{H}+2H^2)(c_{eff}^2-w) +
3H\left[\dot{c_{eff}^2}-(1+c_{eff}^2)\frac{\dot{w}}{1+w}\right]-4\pi
G\bar{\rho}(1+w)(1+3c_{eff}^2)+\frac{k^2c_{eff}^2}{a^2}\right\}\delta&=&0,\,
\end{eqnarray}
\end{widetext}
which is the same result found by \cite{Reis2003}, in the context of
a fluid with EoS parameter $w$. Now, recalling that in our case,
$w=-\frac{\Gamma}{3H}$, we find:
\begin{widetext}
\begin{eqnarray}
\frac{\partial^2\delta}{\partial t^2} + \left[2H + \Gamma + 3c_{eff}^2H - \frac{\Gamma\dot{H}-H\dot{\Gamma}}{H(3H-\Gamma)}\right]\frac{\partial\delta}{\partial t}&+&\nonumber\\
\left\{3(\dot{H}+2H^2)\left(c_{eff}^2+\frac{\Gamma}{3H}\right)
+3H\left[\dot{c_{eff}^2}-(1+c_{eff}^2)\frac{\Gamma\dot{H}-H\dot{\Gamma}}{H(3H-\Gamma)}\right]-4\pi
G\bar{\rho}\left(1-\frac{\Gamma}{3H}\right)(1+3c_{eff}^2)+\frac{k^2c_{eff}^2}{a^2}\right\}\delta&=&0,
\end{eqnarray}
\end{widetext}
or still:
\begin{widetext}
\begin{eqnarray}
\frac{\partial^2\delta}{\partial t^2} + \left[2H + \Gamma + 3c_{eff}^2H - \frac{\Gamma\dot{H}-H\dot{\Gamma}}{H(3H-\Gamma)}\right]\frac{\partial\delta}{\partial t}&+&\nonumber\\
\left\{3H^2\left(c_{eff}^2+\frac{\Gamma}{3H}\right)
+3H\left[\dot{c_{eff}^2}-(1+c_{eff}^2)\frac{\Gamma\dot{H}-H\dot{\Gamma}}{H(3H-\Gamma)}\right]-4\pi
G\bar{\rho}\left[1-\frac{\Gamma^2}{3H^2}+2c_{eff}^2\left(2-\frac{\Gamma}{H}\right)\right]
+ \frac{k^2c_{eff}^2}{a^2}\right\}\delta&=&0. \label{EqDeltaCCDM}
\end{eqnarray}
\end{widetext}
where we have used Eq. (\ref{AccelCCDM}). For numerical purposes it
is interesting to write (\ref{EqDeltaCCDM}) in terms of a new
independent variable, $\eta\equiv{\rm ln}(a(t))$:
\begin{widetext}
\begin{eqnarray}
\label{EqDeltaEta}
\delta'' + \left[2 + 3c_{eff}^2 + \frac{\Gamma + H'}{H} - \frac{\Gamma H'- H\Gamma'}{H(3H-\Gamma)}\right]\delta'&+&\nonumber\\
\left\{3(c_{eff}^2 + c_{eff}^2{}') + \frac{\Gamma}{H}
-3(1+c_{eff}^2)\frac{\Gamma H'- H\Gamma'}{H(3H-\Gamma)}-\frac{4\pi
G\bar{\rho}}{H^2}\left[1-\frac{\Gamma^2}{3H^2}+2c_{eff}^2\left(2-\frac{\Gamma}{H}\right)\right]
+ \frac{k^2c_{eff}^2e^{-2\eta}}{H^2}\right\}\delta&=&0,
\end{eqnarray}
\end{widetext}
where the prime denotes derivative with respect to $\eta$. If we
live in a spatially flat Universe, $K=0$, then $\frac{4\pi
G{\bar \rho}}{H^2}=\frac{3}{2}$, and (\ref{EqDeltaEta}) simplifies to:
\begin{widetext}
\begin{eqnarray}
\label{EqDeltaEtaFlat}
\delta'' + \left[2 + 3c_{eff}^2 + \frac{\Gamma + H'}{H} - \frac{\Gamma H'- H\Gamma'}{H(3H-\Gamma)}\right]\delta'&+&\nonumber\\
+
\left[\left(\frac{\Gamma}{H}-1\right)\left(\frac{\Gamma}{2H}+\frac{3}{2}+3c_{eff}^2\right)+3c_{eff}^2{}'-3(1+c_{eff}^2)\frac{\Gamma
H'- H\Gamma'}{H(3H-\Gamma)} +
\frac{k^2c_{eff}^2e^{-2\eta}}{H^2}\right]\delta&=&0.
\end{eqnarray}
\end{widetext}

Now, if the pressure perturbation vanishes \cite{BasLim},
$c_{eff}^{2}\equiv 0$, then:
\begin{widetext}
\begin{eqnarray}
\label{EqDeltaEtaCef0Flat}
\delta'' + \left[2 + \frac{\Gamma + H'}{H} - \frac{\Gamma H'- H\Gamma'}{H(3H-\Gamma)}\right]\delta'%&+&\nonumber\\
+ \left[\left(\frac{\Gamma}{H}-1\right)\left(\frac{\Gamma}{2H}+\frac{3}{2}\right)-3\frac{\Gamma H'- H\Gamma'}{H(3H-\Gamma)}\right]\delta=0.%&=&0
\end{eqnarray}
\end{widetext}

\section{\label{sec4}Case study: LJO model}
The creation  rate of the LJO model reads (see Eq. (\ref{Gamma}))

\begin{equation}
\label{Gamma1}
\Gamma=3\tilde{\Omega}_{\Lambda}
\left(\frac{\rho_{c0}}{\rho}\right)H,
\end{equation}
whereas the Hubble parameter is given by:
\begin{equation}
H=H_0\left[(\Omega_m-\tilde{\Omega}_{\Lambda}
)a^{-3}+\tilde{\Omega}_{\Lambda} +(1-\Omega_m)a^{-2}\right]^{1/2},
\end{equation}
where $\Omega_m\equiv\frac{\rho_0}{\rho_{c0}}$. In the flat case,
$\Omega_m=1$, and the Hubble parameter reduces to:
\begin{equation}
\label{HaNewCCDM} H=H_0\left[(1-\tilde{\Omega}_{\Lambda})
a^{-3}+\tilde{\Omega}_{\Lambda} \right]^{1/2} \;.
\end{equation}

For the above creation rate, the linear order perturbation equation
(\ref{EqDeltaEta}) can be rewritten as:
\begin{equation}
\delta''+F(\eta)\delta'+G(\eta)\delta=0, \label{npetass}
\end{equation}
where the functions $F(\eta)$ and $G(\eta)$ are given, in the flat
case (\ref{HaNewCCDM}), by:
\begin{widetext}
\begin{eqnarray}
\label{F}
F(\eta) &=& \frac{(1-\tilde{\Omega}_{\Lambda}
  )(1+6c_{eff}^2)+2\tilde{\Omega}_{\Lambda}
e^{3\eta}(8+3c_{eff}^2)}{2(1-\tilde{\Omega}_{\Lambda}+\tilde{\Omega}_{\Lambda} e^{3\eta})} \\
G(\eta) &=& \frac{9(1-\tilde{\Omega}_{\Lambda}
)^2}{2(1-\tilde{\Omega}_{\Lambda} +\tilde{\Omega}_{\Lambda}
e^{3\eta})^2} + \frac{3\tilde{\Omega}_{\Lambda}
e^{3\eta}(5+5c_{eff}^2+c_{eff}^2{}')-3(1-\tilde{\Omega}_{\Lambda}
)(2+c_{eff}^2-c_{eff}^2{}')}{1-\tilde{\Omega}_{\Lambda}
+\tilde{\Omega}_{\Lambda} e^{3\eta}} + \frac{e^\eta
k^2c_{eff}^2}{H_0^2(1-\tilde{\Omega}_{\Lambda}
+\tilde{\Omega}_{\Lambda} e^{3\eta})}. \label{G}
\end{eqnarray}
\end{widetext}

In what follows, the evolution of the density contrast as given
above will be numerically solved by assuming the same initial
conditions of the  Einstein - de Sitter growing model, namely:
$\delta(a_i)=a_i$ and $\delta'(a_i)=1$, where $a_i=10^{-3}$.  It is
natural to impose such conditions because the particle production in
the LJO model is negligible at high redshifts with the model
reducing to the standard dust filled FRW  cosmology. In addition, we
also remark that the integration requires a choice of $c_{eff}^2$
(appearing in the functions $F(\eta)$ and $G(\eta)$) or,
equivalently, the perturbation of the creation pressure.

To begin with, let us recall that $\delta p_c$ is a new degree of
freedom \cite{abramo07,GarrigaMukhanov99} which does not depend
explicitly on the known fluid variables, like $\delta$. Once the
pressure $p_c$ form  is given, one may try to estimate $\delta p_c$.
For the LJO model, by using Eqs. (\ref{Eqp_c}) and (\ref{Gamma1})
one finds:
\begin{equation}
p_c=-\tilde{\Omega}_{\Lambda} \rho_{c0},
\end{equation}
which is time independent. Thus, one could expect $\delta p_c=0$,
which would correspond to $c_{eff}^2=0$. Inserting this condition
into  (\ref{F}) and (\ref{G}) we obtain:
\begin{eqnarray}
F(\eta) &=& \frac{(1-\tilde{\Omega}_{\Lambda} )+16\tilde{\Omega}_{\Lambda} e^{3\eta}}{2(1-\tilde{\Omega}_{\Lambda} +\tilde{\Omega}_{\Lambda} e^{3\eta})}, \\
G(\eta) &=& \frac{9(1-\tilde{\Omega}_{\Lambda}
)^2}{2(1-\tilde{\Omega}_{\Lambda} +\tilde{\Omega}_{\Lambda}
e^{3\eta})^2} + \frac{3(5\tilde{\Omega}_{\Lambda}
e^{3\eta}-2+2\tilde{\Omega}_{\Lambda} )}{1-\tilde{\Omega}_{\Lambda}
+ \tilde{\Omega}_{\Lambda} e^{3\eta}}.
\end{eqnarray}

\begin{figure}[ht]
\includegraphics[width=70mm]{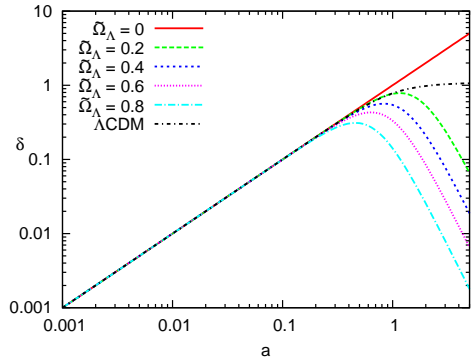}
\caption{$\delta$ as a function of $a$ for the case
 $c_{eff}^2=0$. As discussed in the text, for LJO models 
this is equivalent to the case
$c_{eff}^2=c_s^2$ because the sound speed is zero.  For comparison
we also show the flat
 $\Lambda$CDM case, with $\Omega_m=1-\Omega_{\Lambda}=0.3$ and
$c_{eff}^2=0$.} \label{fig:CCDM0}
\end{figure}

This case is also interesting because the condition for the
Neo-Newtonian perturbed equations to be equivalent to the full
general-relativistic treatment for a single fluid is that
$c_{eff}^2=0$ \cite{Reis2003}.

Another interesting choice to the effective sound speed is
$c_{eff}^2=c_s^2={\dot{p_c}}/{\dot{\rho}}$. In this case, the
non-adiabatic pressure, $\delta p_{nad}\approx\delta
p_c-c_s^2\delta\rho$ vanishes, and one has only adiabatic
perturbations as suggested by CMB observations. However,  the sound
speed of matter in the presence of particle production reads
\begin{equation}
c_s^2 = \frac{H\Gamma'-\Gamma H'-\Gamma(3H-\Gamma)}{3 H(3
H-\Gamma)},
\end{equation}
which vanishes  identically when the creation rate is given by
Eq.(\ref{Gamma1}) since $\dot{p_c}\equiv0$.  Therefore, in the LJO
framework, this choice for the effective sound speed reduces to the
case earlier analyzed ($c_{eff}^2 = 0$). Of course, since
$c_{eff}^2$ is a new degree of freedom, a possibility  is to choose
it as a general time dependent variable,  say, $c_{eff}^2 = w$. The
treatment involving $c_{eff}^2$  as a new degree of freedom is
relevant when the unperturbed fluid equations evolve out of
thermodynamic equilibrium as in the case of LJO model.

In Figure 1, we display the evolution of the density contrast for
LJO model as predicted by the Neo-Newtonian approach. Note that the
increasing mode perturbation grows until its maximum value after
which it decays in the course of the evolution. In particular, for
$\tilde{\Omega}_{\Lambda}=0.2$,  we see that the predicted density
contrast is indistinguishable from that of $\Lambda$CDM cosmology.
However, we stress that such a value is not favored by the
background tests which prefer $\tilde{\Omega}_{\Lambda}=0.7$
\cite{LJO,BasLim}.

In Figure 2, we display the case $c^{2}_{eff}=w = -\Gamma/3H$.  As
shown there, the evolution of the density contrast has been obtained
for different values of ${\tilde \Omega}_{\Lambda}$. Evidently, for
$0.2\le {\tilde \Omega}_{\Lambda} \le 0.4$ the evolution is similar
to the $\Lambda$CDM behavior.

As it appears, the simplest possibility to the quantity $c_{eff}^2$
is to consider it as a constant free parameter and find out which
value is preferred from observations.  In this case,
$c_{eff}^2{}'\equiv 0$ and now we have:
\begin{widetext}
\begin{eqnarray}
\label{FCef} F(\eta) &=& \frac{(1-\tilde{\Omega}_{\Lambda}
)(1+6c_{eff}^2)+2\tilde{\Omega}_{\Lambda} e^{3\eta}(8+3c_{eff}^2)}
{2(1-\tilde{\Omega}_{\Lambda} +\tilde{\Omega}_{\Lambda} e^{3\eta})} \\
G(\eta) &=& \frac{9(1-\tilde{\Omega}_{\Lambda}
)^2}{2(1-\tilde{\Omega}_{\Lambda} +\tilde{\Omega}_{\Lambda}
e^{3\eta})^2} + \frac{15\tilde{\Omega}_{\Lambda}
e^{3\eta}(1+c_{eff}^2)-3(1-\tilde{\Omega}_{\Lambda}
)(2+c_{eff}^2)}{1-\tilde{\Omega}_{\Lambda} +\tilde{\Omega}_{\Lambda}
e^{3\eta}} + \frac{e^\eta
k^2c_{eff}^2}{H_0^2(1-\tilde{\Omega}_{\Lambda}
+\tilde{\Omega}_{\Lambda} e^{3\eta})} \label{GCef}
\end{eqnarray}
\end{widetext}

\begin{figure}[h]
\includegraphics[width=80mm]{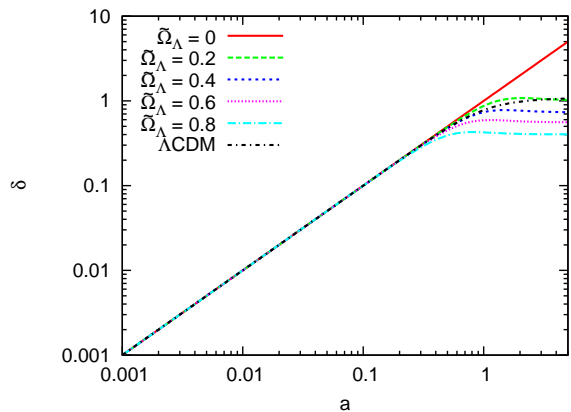}
\caption{$\delta$ as a function of $a$ for the case $c_{eff}^2=w$.
Also shown, for comparison, is the flat $\Lambda$CDM case, with
$\Omega_m=1-\Omega_{\Lambda}=0.3$ and $c_{eff}^2=0$. In this case we
have considered the limit of large scales ($k=0$).}
\label{fig:CCDMW}
\end{figure}

\begin{figure}[h]
\includegraphics[width=80mm]{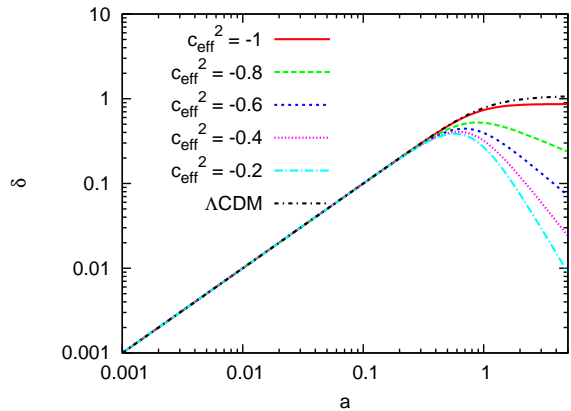}
\caption{$\delta$ as a function of $a$ for the case in which
$c_{eff}^2$ is a free parameter. Also shown, for comparison, is the
flat $\Lambda$CDM case, with $\Omega_m=0.3$ and $c_{eff}^2=0$. In
this case we have considered the limit of large scales ($k=0$), and
$\tilde{\Omega}_\Lambda=0.7$.} \label{fig:CCDMCef}
\end{figure}

In Figure 3, we show the evolution of the density contrast when
$c^{2}_{eff}$ is taken to be a constant free parameter. The selected
values of $c^{2}_{eff}$ are also indicated. Again, we see that there
is a range of values for which the CCDM  matter fluctuation field
mimics the behavior of the concordance $\Lambda$CDM model.

\subsection{The growth rate of clustering}
We would like to end this section with a discussion on the evolution
of the well known indicator of clustering, namely the growth rate
\cite{peeb93}. This is an efficient parametrization of the linear
matter fluctuations $\delta(a)$ which has the following
functional form:
\begin{equation}
\label{fzz221} f(z)=\frac{d{\rm ln}\delta}{d{\rm
ln}a}=-(1+z)\frac{d{\rm ln}\delta}{dz}\;.
\end{equation}

In Table I,  we show the existing growth rate data with the
corresponding error bars, the associated redshifts and related
references.

\begin{table}[ht]
\caption[]{Data of the growth rate of clustering. The correspondence
of the columns is as follows: redshift, observed growth rate and
references.} \tabcolsep 4.5pt
\begin{tabular}{ccc} \hline \hline
z& $f_{obs}$ & Refs. \\ \hline
0.15 & $0.49\pm 0.10$& \cite{Colles01,Guzzo08}\\
0.35 & $0.70\pm 0.18$& \cite{Teg06} \\
0.55 & $0.75\pm 0.18$& \cite{Ross07}\\
0.77 & $0.91\pm 0.36$& \cite{Guzzo08}\\
1.40 & $0.90\pm 0.24$& \cite{daAng08}\\
2.42 & $0.74\pm 0.24$& \cite{Viel04}\\
3.00 & $1.46\pm 0.29$& \cite{McDon05}\\
\end{tabular}
\end{table}

Let us now  attempt to place constraints on the relevant parameters
by performing a standard $\chi^{2}$ minimization procedure between
the observationally measured growth rate and that predicted by the
LJO cosmology. The best fit to the set of independent parameters
($\tilde{\Omega}_{\Lambda}, c_{eff}^{2}$) can be estimated by using
a $\chi^{2}$ statistics with

\begin{equation}
\chi^{2}(\tilde{\Omega}_{\Lambda},c_{eff}^{2})= \sum_{i=1}^{7}
\left[ \frac{f_{obs}(z_{i})- f_{\rm
model}(z_{i},\tilde{\Omega}_{\Lambda},c_{eff}^{2})}
{\sigma_{i}}\right]^{2} \;\;,
\end{equation}
where $\sigma_{i}$ is the observed growth rate uncertainty. Note
that we sample the free parameters as follows:
$\tilde{\Omega}_{\Lambda} \in [0.1,1]$ 
and $c_{eff}^{2} \in [-1.3,2]$ in steps of 0.001.

In Figure 4 (upper panel), we show the 1$\sigma$, 2$\sigma$ and
3$\sigma$
confidence levels in the 
$(\tilde{\Omega}_{\Lambda},c_{eff}^{2})$ plane. It is evident that
$c_{eff}^{2}$ is degenerate with respect to
$\tilde{\Omega}_{\Lambda}$
and that all the values on the interval $-1.3\le c_{eff}^{2} \le 2$
are acceptable within the $1\sigma$ uncertainty. As one may check,
the $c_{eff}^{2}$ parameter as a function of
$\tilde{\Omega}_{\Lambda}$ is well fitted by a power law having the
form:
$$c_{eff}^{2}=
0.333(\pm 0.011)\tilde{\Omega}_{\Lambda}^{-0.989(\pm 0.021)}-1.5
\;.$$

\begin{figure}[ht]
\includegraphics[width=80mm]{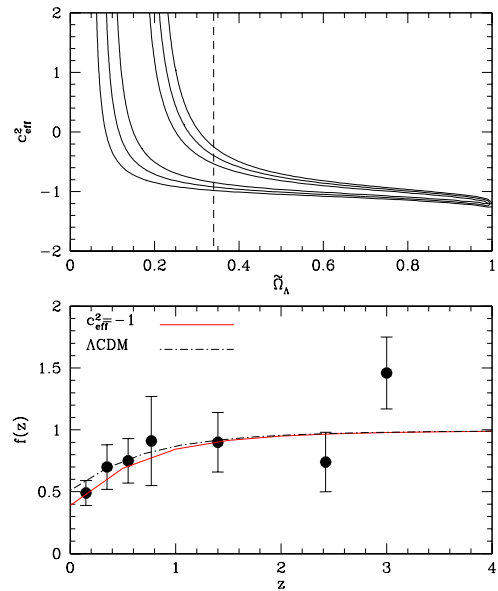}
\caption{{\it Upper Panel:}
Likelihood contours in the 
$(\tilde{\Omega}_{\Lambda},c_{eff}^{2})$ plane. The contours are
plotted where $-2{\rm ln}({\cal L/L_{\rm max}})$ is equal to 2.32,
6.16 and 11.83, corresponding to $1\sigma$, $2\sigma$ and $3\sigma$
confidence level. The perpendicular dashed line separates the
regions in which the flat LJO model can accommodate an accelerated
expansion ($\tilde{\Omega}_{\Lambda}>1/3$), 
equivalent to the standard $\Lambda$CDM  cosmology. {\it Bottom
Panel:} Comparison of the observed (solid circles see Table I) and
theoretical evolution of the growth rate of clustering $f(z)$. Note
that we have imposed
$\tilde{\Omega}_{\Lambda}=0.7$ as suggested by the background tests \cite{LJO,BasLim}. 
The lines correspond to the LJO (solid curve with $c_{eff}^{2}=-1$)
and the $\Lambda$CDM (dot-dashed curve with $\Omega_{\Lambda}\equiv
\tilde{\Omega}_{\Lambda}$ and $c_{eff}^{2}\equiv 0$) models.}
\end{figure}
It is also evident that the likelihood analysis puts some
constraints on the value of $c_{eff}^{2}$ once including the
necessary condition for an inflection point, that is, a transition
from a decelerating to an accelerating regime
in our past. Thus, for $\tilde{\Omega}_{\Lambda}> 1/3$, 
(see dashed line in the upper panel of figure 4) we find that
$c_{eff}^{2}$ lies in the interval
$-1.24\le c_{eff}^{2} \le -0.72$. 
In this framework, if we marginalize
over $\tilde{\Omega}_{\Lambda}=0.7$, 
then the likelihood analysis provides a best fit value of
$c_{eff}^{2} \simeq -1$ with $\chi^{2}\simeq 4.03$ (${\rm dof}=6$).
The latter result remains unaltered within a physical range of $0.65
\le \tilde{\Omega}_{\Lambda} \le 0.74$.
Of course, for comparison we perform the same analysis also for the
$\Lambda$CDM model ($\Omega_{\Lambda}=0.7$) and we find
$\chi^{2}\simeq 4.88$ (${\rm dof}=7$).

In the bottom panel of  Figure 4, we present  the LJO growth rate of
clustering (solid line). We find that for $z\le 1$ the LJO growth
rate is somewhat less with respect to that of the usual $\Lambda$
cosmology (dot-dashed line) and such a difference increases for
extremely low redshifts. However, at relatively large redshifts
($z>1$) the growth rate of the LJO model is indistinguishable from
that of the $\Lambda$CDM model. This should be expected because the
flat LJO model reduces to the Einstein - de Sitter cosmology at
intermediate and high $z$'s (negligible particle production). Note
that the measured $f_{obs}(z)$ are presented in the bottom panel of
Fig. 4 by the filled symbols.

We further explore our statistical results by using a Bayesian
statistics (see for example \cite{essence}), in which the
corresponding estimator is defined as: $BIC=\chi^{2}+k_{n}{\rm ln}N$
[where $k_{n}$ is the number of free parameters and $N(\equiv 7)$ is the
number of data points used in the fit]. The next step is to estimate
the relative deviation between the two models $\Delta
BIC=BIC_{LJO}-BIC_{\Lambda}$. In general a difference in $BIC$ of
$\Delta BIC>6$, is considered evidence against that model which
occurs the larger $BIC$. In our case, we find $\Delta BIC \simeq
1.01$ which implies that the LJO model with
$(\tilde{\Omega}_{\Lambda},c_{eff}^{2})=(0.70,-1)$ fits
very well the growth rate data. The latter result holds also for
$0.65 \le \tilde{\Omega}_{\Lambda} \le 0.74$.

Secondly, we compare the growth rate of clustering between data and
models via a Kolmogorov-Smirnov (KS) statistical test respectively
by computing the corresponding consistency between models and data
(${\cal P}_{KS}$). Although both cosmological models fit well the
data it seems that the LJO model provides a slightly better fit
(${\cal   P}_{KS}\simeq 0.997$) with respect to that of the usual
$\Lambda$ cosmology (${\cal P}_{KS}\simeq 0.883$). Note that the KS
test between the two cosmological models gives ${\cal P}_{KS}\simeq
1$.

Based on the above statistical tests it becomes evident that in the
LJO model ($0.65 \le \tilde{\Omega}_{\Lambda} \le 0.74$,
$c_{eff}^{2}=-1$), the corresponding Hubble flow as well as the
matter fluctuation field resembles that of the traditional
$\Lambda$CDM cosmology  without the need of the required, in the
classical cosmological models, dark energy.

At this point one may ask about  the possibility of future 
detection or at least what is a clear cosmic 
signature of this gravitationally induced creation process of 
cold dark matter particles.  
As it is widely known, some very massive dark matter particle 
candidates like the wimpzillas ($M_W \sim10^{13}$ GeV)  
can be copiously produced only at the very early stages of 
the Universe, mainly at the end of inflation \cite{KST,PDM}, and, as such, this 
kind of candidate does not 
fit in our phenomenological treatment for continuous matter creation. 
In principle, a 
more realistic scenario is provided by  the semi-classical approach proposed 
by Parker and collaborators \cite{Parker}, where massive particles 
(associated to a real scalar field) can be continuously 
created during the expansion of the Universe. On the other hand, since 
the current dark matter detection 
experiments rely on the physical properties 
of the dark matter (mass, cross section, etc), we are 
unable to identify  (based only on our phenomenological 
approach) which is the specific candidate for the continuous 
cold dark matter production discussed here. However, 
by assuming that the
created mass are of the form of neutralinos ($M_N \sim 100 GeV$), one may show 
that its present creation rate,  
$\Gamma_{neu}\sim 10^{-11}$ cm$^{-3}$yr$^{-1}$,  has not 
changed appreciably in the last few billion years when the 
Universe entered its accelerating phase \cite{LJO}.

\section{Conclusions}

The problem related to the nature of non-baryonic component filling the observed Universe is
usually referred to as the non-baryonic dark 
matter problem. In the last three decades, many 
possible candidates from particle physics have been proposed 
to describe such a dark matter component \cite{PDM}. 
In the framework of general relativity, it is also interesting to know whether the dark sector (dark
matter + dark energy) could be properly reduced
only to the dark matter component with the conventional $\Lambda$CDM emerging as an effective 
description \cite{LJO,BasLim}. Naturally, unlike the decelerating dust-filled Einstein-de Sitter Universe, 
an extra mechanism must be responsible for the present observed  accelerating stage.

In this context, we have performed a Neo-Newtonian description of relativistic CCDM
models, that is, models endowed with gravitationally induced
production of cold dark matter particles. The complete equivalence
of the Neo-Newtonian cosmological approach with the general
relativistic background equations regardless of the specific form
adopted to the particle production rate, $\Gamma$, has been
established. In the same vein, the general perturbed equations has
also been derived in this framework. Due to the form of the negative
creation pressure ($p_c = -\Gamma \rho/3H$) we have determined the
general differential equation to the contrast density by assuming a
full equivalence to a fluid model with an EoS parameter
$w=-{\Gamma}/{3H}$. The resulting equations of section IV are also
valid  for arbitrary forms of $\Gamma$.

In section V, we have focused our attention to a specific CCDM
cosmology, namely, the LJO model \cite{LJO}. The interest for this
special class of CCDM cosmology comes from the fact that its
expanding history is equivalent to the standard $\Lambda$CDM model.

Three different perturbed scenarios depending on the choice of the
$c_{eff}^{2}$ were analyzed. In the first one
$c_{eff}^{2}=c_s^{2}=0$ and the  linear perturbation is equivalent
to that predicted by $\Lambda$CDM only up to redshifts of the order
of one. It was found that the increasing mode decays rapidly after
$z \sim 1$ with the redshift marking the beginning of attenuation
depending on the $\tilde{\Omega}_{\Lambda}$ parameter (see Fig. 1).
We also show that the nonadiabatic case with $c_{eff}^{2}=\omega$ is
equivalent to a perturbed $\Lambda$CDM, however, only for $0.2\le
{\tilde \Omega}_{\Lambda} \le 0.4$ (see Fig. 2). Finally,  when
$c_{eff}^2$ was treated as a free constant parameter, the perturbed
evolution is like $\Lambda$CDM  for $\tilde{\Omega}_{\Lambda}=0.7$
(see Fig. 3) which is  exactly the same value preferred by the
expanding cosmic history \cite{LJO}. For all these models, it should
be interesting to obtain the predicted matter power spectrum in the
context of the relativistic formulation. This work is in progress
and will be published elsewhere.

For completeness, we also performed a detailed statistical analysis
based on the observed growth rate of clustering in order to
constrain the free parameters ($\tilde{\Omega}_{\Lambda},
c_{eff}^{2}$). Interestingly, when combined with the background
tests \cite{LJO,BasLim}, the best fit values are
$\tilde{\Omega}_{\Lambda} = 0.7$ and $c_{eff}^{2} = -1$ (compare
Figs. 3 and 4). In this case, the LJO pattern predicts an overall
dynamics (Hubble flow and matter fluctuation field) which is for all
practical purposes indistinguishable from the traditional $\Lambda$
cosmology. Naturally, such a solution based on a reduction of the
dark sector, provides not only a possible reinterpretation of the
$\Lambda$CDM cosmology but also offers a viable alternative cosmic
scenario to the late time accelerating Universe without the need of
an exotic dark energy. Indeed one of the main advantages of such
CCDM cosmology is the fact that it contains the same number of free
parameters as the concordance $\Lambda$CDM model, and, therefore, it
does not require the introduction of any extra fields in its
dynamics.

\end{document}